\documentclass[journal]{IEEEtran}

\usepackage{xcolor} 
\usepackage{graphicx}
\usepackage{color}

\usepackage{amsmath}
\usepackage{amssymb}
\usepackage{amsfonts}

\usepackage{caption}
\usepackage{subcaption}

\usepackage{float}
\usepackage{url}
\usepackage{hyperref}
\hypersetup{
    colorlinks = true,
    linkcolor = black,
    citecolor = black,
    citebordercolor = black,
    urlcolor = black,
    urlbordercolor = black,
}

\usepackage{booktabs}
\usepackage{multirow}
\usepackage{ulem}
\usepackage{cite}

\newcommand{\loss}{\mathcal{L}}

\newcommand{\bk}{\phi}
\newcommand{\bh}{\boldsymbol{h}}
\newcommand{\bH}{\boldsymbol{H}}
\newcommand{\bx}{\boldsymbol{x}}
\newcommand{\bz}{\boldsymbol{z}}
\newcommand{\by}{\boldsymbol{y}}

\newcommand{\bC}{\boldsymbol{C}}
\newcommand{\predv}{\text{Pred}^V}
\newcommand{\predb}{\text{Pred}^B}

\newcommand{\bX}{\boldsymbol{X}}
\newcommand{\bc}{\boldsymbol{c}}

\newcommand{\nhis}{N_{\text{his}}}
\newcommand{\nhistrain}{N_\text{his}^\text{train}}
\newcommand{\nhisdecode}{N_\text{his}^\text{decode}}
\newcommand{\bo}{\boldsymbol{o}}

\definecolor{ogreen}{HTML}{3C8031}
\definecolor{oyellow}{HTML}{FFCC99}

\usepackage[]{ulem}
\newcommand\st{\bgroup\markoverwith{\textcolor{red}{\rule[0.5ex]{2pt}{0.8pt}}}\ULon}

\usepackage[sort&compress,numbers]{natbib}

\begin{document}

\title{Advanced {Long-Content} Speech Recognition with Factorized Neural Transducer}

\author{
Xun Gong, \IEEEmembership{Student Member,~IEEE,}
Yu Wu, \IEEEmembership{Member,~IEEE,}
Jinyu Li, \IEEEmembership{Senior Member,~IEEE,}
Shujie Liu, \IEEEmembership{Member,~IEEE,}
Rui Zhao, \IEEEmembership{Member,~IEEE,}
Xie Chen, \IEEEmembership{Member,~IEEE,}
and Yanmin Qian, \IEEEmembership{Senior Member,~IEEE}
\thanks{


This work was supported in part by China NSFC Projects under Grants 62122050 and 62071288, and in part by Shanghai Municipal Science and Technology Commission Project under Grant 2021SHZDZX0102. (Corresponding author: Yanmin Qian.)

Xun Gong, Xie Chen and Yanmin Qian are with the Auditory Cognition and Computational Acoustics Lab, the Department of Computer Science and Engineering and the MoE Key Laboratory of Artificial Intelligence, AI Institute, Shanghai Jiao Tong University, Shanghai 200240, China (e-mail: gongxun@sjtu.edu.cn; chenxie95@sjtu.edu.cn; yanminqian@sjtu.edu.cn).

Yu Wu, Jinyu Li, Shujie Liu and Rui Zhao are with Microsoft Corporation (e-mail: wumark@126.com; jinyli@microsoft.com; shujliu@microsoft.com; ruzhao@microsoft.com). 
}
}

\markboth{IEEE/ACM TRANSACTIONS ON AUDIO, SPEECH, AND LANGUAGE PROCESSING, VOL. 32, 2024}%
{Shell \MakeLowercase{\textit{et al.}}: Bare Demo of IEEEtran.cls for IEEE Journals}

\maketitle

\begin{abstract}
Long-content automatic speech recognition (ASR) has obtained increasing interest in recent years, as it captures the relationship among consecutive historical utterances while decoding the current utterance.
In this paper, we propose two novel approaches, which integrate {long-content} information into the factorized neural transducer~(FNT) based architecture in both non-streaming~(referred to as \textit{LongFNT}) and streaming~(referred to as \textit{SLongFNT}) scenarios.
We first investigate whether {long-content} transcriptions can improve the vanilla conformer transducer~(C-T) models.
Our experiments indicate that the vanilla C-T models do not exhibit improved performance when utilizing {long-content} transcriptions, possibly due to the predictor network of C-T models not functioning as a pure language model. 
Instead, FNT shows its potential in utilizing {long-content} information, where we propose the \textit{LongFNT} model and explore the impact of {long-content} information in both text~(LongFNT-Text) and speech~(LongFNT-Speech).
The proposed LongFNT-Text and LongFNT-Speech models further complement each other to achieve better performance, with transcription history proving more valuable to the model.
The effectiveness of our LongFNT approach is evaluated on LibriSpeech and GigaSpeech corpora, and obtains relative 19\% and 12\% word error rate reduction, respectively.
Furthermore, we extend the LongFNT model to the streaming scenario, which is named \textit{SLongFNT}, consisting of SLongFNT-Text and SLongFNT-Speech approaches to utilize {long-content} text and speech information.
Experiments show that the proposed SLongFNT model achieves relative 26\% and 17\% WER reduction on LibriSpeech and GigaSpeech respectively while keeping a good latency, compared to the FNT baseline.
Overall, our proposed \textit{LongFNT} and \textit{SLongFNT} highlight the significance of considering {long-content} speech and transcription knowledge for improving both non-streaming and streaming speech recognition systems.
\end{abstract}

\section{Introduction}
\label{sec:intro}

\IEEEPARstart{E}{nd}-to-end~(E2E) automatic speech recognition~(ASR) models \cite{E2EOverview, prabhavalkar2023end}, including connectionist temporal classification~(CTC)~\cite{graves2006ctc}, attention-based encoder-decoder~(AED)~\cite{vaswani2017attention,hori2017joint,karita2019comparative,watanabe2017hybrid,chan2016listen,chorowski2015attention,Chiu2017StateoftheArtSR,gong2022kd,gong2021accent,qian2022accent}, and recurrent neural network transducer~(RNN-T)~\cite{graves2012sequence,chen2021developing,Sainath2020ASO,Battenberg2017ExploringNT,Li2020OnTC} have become the dominated models, surpassing traditional hybrid models~\cite{Sainath2020ASO, Li2020Developing}.
A common practice for ASR is to train the model with individual utterances without considering the correlation between utterances.
{In real-world scenarios like conversations and meetings, speech often appears in long-content formats. This context-rich nature provides an opportunity to enhance recognition accuracy compared to isolated short utterances.}
For example, certain keywords mentioned earlier may reappear later in a dialogue, or the same acoustic environment can be used to guide the recognition in the future.
\textbf{{Long-content} ASR} (also conversational ASR, dialog-aware ASR or large-context ASR), is a special version of the ASR task that aims to improve ASR accuracy by {capturing the relationships between the current decoded utterance and consecutive historical utterances pre-segmented by voice-activity-detection~(VAD)}.

In AED architecture, previous approaches to model {long-content} scenarios mainly includes concatenating consecutive speech or transcriptions of utterances~\cite{kim2018dialogcontext,hori2020transformerbased} and using auxiliary encoders to model context information in an AED manner \cite{masumura2019largea, masumura2021hierarchicala}.
Hori et al.~\cite{hori2021advanced} extended their context-expanded transformer to accelerate the decoding process in streaming AED architecture.
Recurrent neural language models \cite{irie2019trainingb, mikolov2012contexta,tran2016inter,mikolov2012context,liu2017dialog,wang2016larger,lin2015hierarchical,chiu2021crosssentence,xiong2018session,gong2023faed} can also be used with consecutive {long-content} transcriptions.
Recently, Wei et al. \cite{wei2022conversational, wei2022leveraging, wei2022improving} proposed using a latent variational module, context-aware residual attention, and pre-trained encoders to leverage acoustic and text content. These methods offer more comprehensive approaches to improve ASR performance by capturing {long-content} information.

Transducer-based systems, such as recurrent neural transducer (RNN-T), transformer transducer (T-T) are becoming more popular in industry due to their natural streaming capabilities and low latency, as well as their perceived robustness compared to attention-based systems \cite{chiu2019comparison,rao2017exploringa,He2018StreamingES,E2EOverview}.
Narayanan et al. \cite{narayanan2019recognizing} had conducted primitive explorations by simulating {long-content} training and adaptation to improve performance using short utterances.
Schwarz et al. \cite{schwarz2021improving} showed that combining input and context audio helps the network learn both speaker and environment adaptations.
Kojima \cite{kojima2021largecontext} explored the utilization of large context.
However, incorporating consecutive transcription history into the neural transducer model is still an open area that has not been well explored.
{
In modern ASR systems, however, streaming ASR shows its importance on reducing runtime complexity and real-time efficiency~\cite{zeyer2023monotonic,moritz2019streaming,shi2021emformer}.
But there is usually a performance degradation compared with the offline systems which is due to the lack usage of history information.
Few works besides Hori et al. \cite{hori2021advanced} have explored {long-content} ASR to resolve these problem in streaming AED situation.
Hence there is still much needed to be done to fully realize the potential of {long-content} neural transducer models in streaming ASR applications.
}

In this paper, we propose the novel non-streaming and streaming transducer models, \textit{LongFNT} and \textit{SLongFNT}, respectively, which integrate {long-content} information into the factorized neural transducer (FNT)~\cite{chen2022factorized,zhaorui,gong2022longfnt} architecture to solve the above challenge.

We firstly attempted to embed {long-content} transcriptions into the predictor of the vanilla neural transducer in non-streaming situation, but our experiments showed that it had limited impact on the performance.
One possible explanation for the limited impact of {long-content} transcriptions in the vanilla neural transducer is that the prediction network does not function as a pure language model (LM), which constrains its ability to model {long-content} transcriptions.
It also indicates that effective methods for AED models, such as those proposed in {Hori et al. \cite{hori2021advanced}}, cannot be extended to transducer-based models, as they rely heavily on the LM characteristic of the decoder.
Then, we utilized the {FNT (or named as modular hybrid autoregressive transducer)} architecture \cite{chen2022factorized,variani2020hybrid,zhaorui,zeyer2020new,meng2022modularhat}, which factorizes the blank and vocabulary prediction modules, allowing for the use of a standalone LM for vocabulary prediction.

Based on FNT, we propose the \textbf{LongFNT} architecture, by fusing two sub-architecture, LongFNT-Text and LongFNT-Speech to utilize {long-content} text and speech individually.
This architecture is proposed in our recent publication~\cite{gong2022longfnt}.
For LongFNT-Text, utterance-level integration and token-level integration are proposed to integrate high-level {long-content} features from historical transcriptions based on the vocabulary predictor.
Concretely, a context encoder is employed to yield the embedding of each token in historical transcriptions.
To further improve performance, we employed a pre-trained text encoder, RoBERTa~\cite{roberta2020}.
Besides, we embed {long-content} speech into the encoder to propose the LongFNT-Speech model.

Furthermore, we propose \textbf{SLongFNT} model, which extends our non-streaming LongFNT model into the streaming scenario.
Similarly two approaches SLongFNT-Text and SLongFNT-Speech are proposed.
The SLongFNT-Text uses LSTM~\cite{Hochreiter1997LongSM} as the vocabulary predictor backbone and traditional attention to integrate {long-content} information at the token level.
Additionally, we explored the use the vocabulary predictor's hidden state as context to reduce computational cost.
For real-time speech processing, we developed SLongFNT-Speech, which uses {long-content chunk-based} attention and integrates historical utterances with the hidden layer representation of the current chunk for key and value.
We explored several ways to downsample the historical features, including statistical and dilated downsampling.

The main contributions of this paper can be summarized as follows:
\begin{itemize}
    \item
    {Firstly, we introduce the concept of long-content scenarios in ASR, as an opportunity to utilize both historical audio and textual information.}
    
    \item For non-streaming {long-content} scenario, we proposed two architectures to explore the {long-content} information from {historical content} text and speech respectively, i.e. LongFNT-Text and LongFNT-Speech, and further combined these two methods into one integrated architecture to obtain the final \textbf{LongFNT} model, which {is built upon our preliminary work~\cite{gong2022longfnt} with deeper analysis}.

    \item For streaming {long-content} scenario, we further improved the structure of LongFNT to meet the real-time requirements under streaming conditions. 
    Similarly, we propose SLongFNT-Text and SLongFNT-Speech, and finally combine them to obtain the \textbf{SLongFNT} model.
    Huge accuracy improvement and low latency increase also can be observed in this streaming scenario.
\end{itemize}

The rest of the paper is organized as follows. Neural transducer architectures are first reviewed in Section~\ref{sec:related}. Section~\ref{sec:non-streaming_long_form} presents the newly proposed LongFNT which utilizing the {long-content} information in ASR, and Section~\ref{sec:streaming_longform} further explored the streaming version named SLongFNT.
The detailed experimental setup, results and analysis are described in Section~\ref{sec:expr_setup},\ref{sec:expr}, and finally the conclusions are given in Section~\ref{sec:concl}.

\section{Revisit on Neural Transducer}
\label{sec:related}

\subsection{Transformer-based Neural Transducers}
\label{sec:transducer}

{Take the acoustic features $\bX = \{\bx_1, \cdots, \bx_T\}$ as input and label sequence $\by = \{y_1, \cdots, y_L\}$ as output,} modern transducer architecture for speech recognition consists of 3 components, a {speech encoder}, a label predictor and a joint network to predict the tokens.
Conformer~\cite{gulati2020conformer} is a convolutional augmented transformer speech encoder that is widely used in attention-based encoder-decoder and neural transducer architectures to improve the ASR performance.
The predictor works like a language model which produces label representation $\bz_l$ given non-blank outputs $\by_{\le l}$, where $t$ is the time index and $l$ is the output label index.
The encoder and predictor outputs are combined in the joint network and are subsequently passed through the output layer to compute the probability distribution $\bz_{t,l}$ over the output layer:
\begin{align}
  \bh_{{[1:T]}} &= \text{Encoder} ({\bX}), \label{eq:enc} \\
  \bz_l &= \text{Predictor} (\by_{\le l}), \label{eq:dec} \\
  \bz_{t,l} &= \text{JointNet} (\bh_t, \bz_l) \label{eq:joint}
\end{align}
where $t \in [1, T], l \in [1, L]$ are the frame and label index, respectively.
The predicted probability of the neural transducer model and loss can be computed as:
\begin{align}
  P_{ASR}(\hat{y}_{l+1} | \bx_{\le t}, y_l) &= \text{softmax} (\bz_{t,l}), \\
  \loss_{transducer} &= - \log \sum_{\alpha \in \eta^{-1} (\by)} P(\alpha|\bX), \label{eq:loss_rnnt}
\end{align}
where $\eta$ is a many-to-one function from all possible transducer paths to the target $\by$ and a special blank symbol, $\bk$, is added to the output vocabulary.
Therefore the output set is $\{\bk \cup \mathcal{V}\}$ , where $\mathcal{V}$ is the vocabulary set.

\subsection{Factorized Neural Transducer}
\label{sec:fnt}

\begin{figure}[ht]
  \centering
  \includegraphics[width=\linewidth]{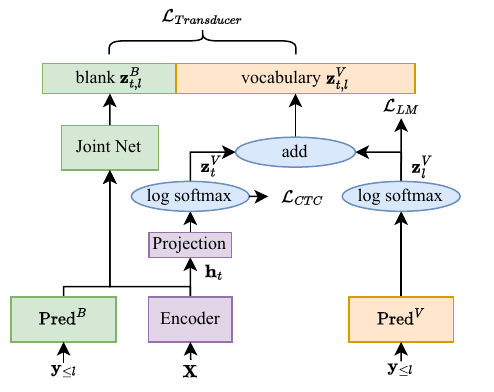}
  \caption{Illustration of factorized neural transducer~(FNT) and its improved version~\cite{zhaorui}.}
  \label{fig:fnt}
\end{figure}

The factorized neural transducer model~(FNT)~\cite{chen2022factorized,zhaorui} has emerged as a promising architecture, aiming to separately predict the blank token and vocabulary tokens, so that an auxiliary vocabulary predictor fully functions as an LM.
The FNT model is comprised of four main components, the conformer encoder, the blank predictor~($\predb$), the joint network for $\bk$, and the vocabulary predictor~($\predv$, i.e. LM manner).
The whole architecture is illustrated in Figure~\ref{fig:fnt}.
\begin{align}
  \bz_{t,l}^B &= \text{Joint} (\bh_t, \predb (\by_{\le l})), \label{eq:predb} \\
  \bz_{l}^V &= \text{log\_softmax} (\predv (\by_{\le l})) \nonumber \\
       &= \log P_{LM}(\hat{y}_{l+1} | \by_{\le l}), \label{eq:predv_v} \\
  \bz_{t}^V &= \text{log\_softmax} (\text{Linear} (\bh_t)), \label{eq:predv_t} \\
  \bz_{t,l}^V &= \bz_{t}^V\text{[:-1]} + \beta \bz_{l}^V, \label{eq:predv}
\end{align}
where $\beta$ is a trainable parameter, $\bz_{t}^V$ is used for computing CTC loss $\loss_{CTC}$, and $\bz_t^V$ become $U+1$ because CTC needs another extra output blank $\psi$.
{It's important to note that the blank token ($\psi$) in CTC is not the same as the blank token ($\phi$) in RNN-T mentioned above~\cite{graves2006ctc,graves2012sequence}.}
For $\predb$, it repeats Equation~\ref{eq:dec},\ref{eq:joint} but only predict the $\bk$ in Equation~\ref{eq:predb}.
For $\predv$, LM logits is firstly projected to the vocabulary size and converted to the log probability domain by the log softmax operation, denoted as $\bz_{l}^V$ in Equation~\ref{eq:predv_v}.
In Equation~\ref{eq:predv}, then it is added with projected encoder output $\bz_{t}^V$ to get $\bz_{t,l}^V$, i.e. the output of $\predv$.
Finally we can compute the posterior of the transducer model as
\begin{align}
  P_{ASR}(\hat{y}_{l+1} | \bX, y_l) = \text{softmax} ([\bz_{t,l}^B, \bz_{t,l}^V]).
\end{align}
and then the loss is computed as
\begin{align}
  \loss = \loss_{transducer} + \lambda_{LM} \loss_{LM} + \lambda_{CTC} \loss_{CTC},
\end{align}
where $\lambda_{LM}, \lambda_{CTC}$ are hyper-parameters.

\section{LongFNT: {Long-content} Factorized Neural Transducer ASR}

\label{sec:non-streaming_long_form}

In this section we describe LongFNT, the non-streaming {long-content} ASR architecture based on factorized neural transducer, which contains LongFNT-Text and LongFNT-Speech parts.

Here, we use $p$ as the index of the current utterance, and then the first/second utterance before utterance $p$ are $p-1$ and $p-2$ and etc.
{
Then the historical utterances have acoustic sequences like $\{\cdots, \bX^{p-2}, \bX^{p-1}\}$. and label sequences like $\{\cdots, \by^{p-2}, \by^{p-1}\}$, where the current utterance has acoustic sequence $\bX^p$ and label sequence $\by^p$ and $\bX^p = [\bx^p_1, \cdots, \bx^p_{T^p}], \by^p = [y^p_1, \cdots, y^p_{L^p}]$.
And then the acoustic representation sequence from speech encoder is $\bh^p_{[1:T^p]}$ for currernt utterance $p$.
}

\subsection{LongFNT-Text: {Long-content} Text Integration of FNT}
\label{sec:longfnt_text}

\begin{figure}[ht]
  \centering
  \includegraphics[width=\linewidth]{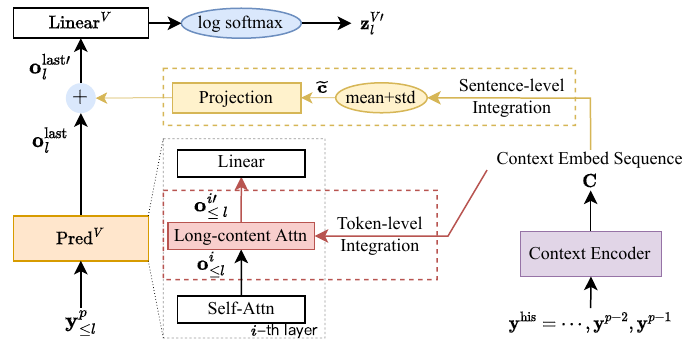}
  \caption{Architecture of LongFNT-Text: the text-side context encoder and {long-content} textual integration methods for $\predv$}
  \label{fig:longfnt-text}
\end{figure}

As shown in Figure~\ref{fig:longfnt-text}, we first propose LongFNT-Text, which contains a text-side context encoder and two different textual integration methods.
We modify the $\predv$ in Equation~\ref{eq:predv_v} to expand {long-content} textual information in FNT.

\textbf{Context Encoder}: 
The text-side context encoder converts historical label sequences $\by^{\text{his}} = \{\cdots, \by^{p-2}, \by^{p-1}\}$ into token-level historical embedding sequence $\bC$:
\begin{align}
  \bC = \text{Context-Encoder} (\by^{\text{his}}), \label{eq:context}
\end{align}
where $\bC$ has length equal to $\cdots + L^{p-2} + L^{p-1}$.
During inference, the current utterance information will not be used in the context encoder.
The basic context encoder is jointly trained with {long-content} FNT in a transformer manner.
To extract stronger features, we directly use a pre-trained RoBERTa~\cite{roberta2020} model as the context encoder.
{We utilize the pre-trained RoBERTa}\footnote{https://huggingface.co/sentence-transformers/all-roberta-large-v1} {model and freeze it during LongFNT training.}

Two textual integration methods are designed for LongFNT-Text.
As shown before, FNT factorizes out the vocabulary predictor part by jointly training the speech-text pair data, therefore the historical transcriptions can be injected inside the vocabulary predictor $\predv$ or after it.

\textbf{{Utterance-level} integration}:
{To get the utterance-level embedding $\tilde{\bc}$}, we first do mean and standard variance (std) such that $\tilde{\bc} = \text{concat} (\text{mean}(\bC), \text{std} (\bC))$, {{and then enhance $\bz^V_l$ with utterance-level information $\tilde{\bc}$} in the yellow box of Figure~\ref{fig:longfnt-text}}:
\begin{align}
  \bo_{\le l}^{\text{last}} &= 
  {\text{$\predv$-Encoder}} (\by_{\le l}), \\
  \bo_l^{\text{last}'} &= \bo_l^{\text{last}} {+} \text{Projection} (\tilde{\bc}), \\
  \bz^{V'}_l &= \text{log\_softmax} (\text{Linear}^V \cdot \text{ReLU} (\bo_l^{\text{last}'})),
\end{align}
where $\bo_{\le l}^{\text{last}}$ is the $l$-th textual embedding {in the last layer of $\predv$-Encoder}, and $\bo_l^{\text{last}'}$ is the {historical-enhanced} version of $\bo_l^{\text{last}}$, and $\bz^{V'}_l$ is the {historical-enhanced} logits. 

\textbf{Token-level integration}:
Different from the utterance-level integration method, we put more granular historical information ($\bC$) into $\predv$ by adding an auxiliary cross-attention layer inside transformer blocks:
\begin{align}
  \bo^i_{\le l} &= \text{MHA} (\bo^{i-1}_{\le l}, \bo^{i-1}_{\le l}, \bo^{i-1}_{\le l}), \\
  \bo^{i'}_{\le l} &= \text{MHA} (\bo^i_{\le l}, \bC, \bC), \label{eq:fnt_text} \\
  \bo^{i'}_{\le l} &= \text{FFN} (\bo^{i'}_{\le l}),
\end{align}
where $\predv$ is a transformer encoder, $i$ is the block index, $\bo_{\le l}$ is the representation of current tokens, FFN is the feed-forward layer and MHA is the multi-head attention layer, respectively. Residual connection is ignored for simplification.
In Equation~\ref{eq:fnt_text}, we integrate the {long-content} {context} embedding sequence $\bC$ into the representation $\bo_{\le l}$, where $\bC$ is key and value and $\bo^i_{\le l}$ is the query in this attention.
Meanwhile, there is also a projection layer for $\bC$ when we use mismatched dimensions for the context encoder and $\predv$.

Finally, the utterance-level integration and the token-level integration can be combined to achieve better utilization of $\bC$.
Since the $\predv$ in FNT is designed to be an LM, we explore the possibility of training it independently on a much larger text corpus {(i.e. the external text)} than the transcriptions in the FNT training data.
The vocabulary of the external LM is the same as that of the FNT system, and is {pre-trained} using the conventional cross-entropy loss. 
With the help of large external text data, the model achieves better results, which is then named as \textbf{LongFNT-Text}.

\subsection{LongFNT-Speech: {Long-content} Enhanced Speech Encoder}
\label{sec:longfnt_speech}

\begin{figure}[ht]
  \centering
  \includegraphics[width=0.9\linewidth]{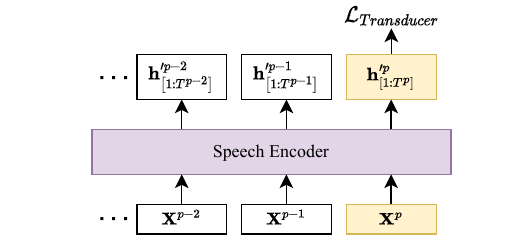}
  \caption{Architecture of LongFNT-Speech: the speech encoder pipeline with {long-content} speech input. Only the yellow part is used for gradient back propagation.}
  \label{fig:longfnt-speech}
\end{figure}

As shown in Figure~\ref{fig:longfnt-speech}, we describe our proposed LongFNT-Speech to train the {speech encoder} on {long-content} speech, which is pre-segmented into utterances $\{\cdots, \bX^{p-2}, \bX^{p-1}\}$.
\begin{align}
  \cdots, \bh^{'p-1}_{[1:T^{p-1}]}, \bh^{'p}_{[1:T^p]} = \text{Encoder} (\cdots, \bX^{p-1}, \bX^{p}), 
\end{align}
where $\bh^{'p}_{t}$ is the {historical-enhanced} $t$-th acoustic representation of the $p$-th utterance~(i.e. the current one) matches $\bh_t$ in Equation~\ref{eq:enc}.
The label representations $\bz_t^V$ is calculated by $\bh^{'p}_t$ as in Equation~\ref{eq:predv}.
{For normal FNT, the whole $\cdots, \bh^{'p-1}, \bh^{'p}_{\{1:T\}}$ is used to compute the transducer loss and for gradient back-propagation.
However, LongFNT-Speech only utilizes the current acoustic hidden representations $\bh^{'p}_{[1:T]}$ for computing the transducer loss and CTC loss, and backward across historical utterances is ignored during training.
}
Using such extension, the speech encoder receives a longer history and thus benefits both training and evaluation.

\subsection{Training strategies for LongFNT}
\label{sec:longfnt_train}

Combining LongFNT-Text and LongFNT-Speech, the final proposed method is called as \textbf{LongFNT}.
During training, the large-context encoder obtains not hypotheses but reference~(oracle) transcripts.
The {long-content} information utilization is controlled by the number of historical utterances~($\nhis$), and will effect the improvement of the LongFNT model. 
$\nhis$ is a hyper-parameter, and $\nhistrain$ (the number of historical utterances during training) is randomly sampled from the {pre-defined} distribution $[0, \text{$\nhis$}]$ for each utterance, {while $\nhisdecode$ (the number of historical utterances during decoding) is controlled by $\nhis$ and available historical utterances.
Take $\nhis=3$ as an example, the first decoded utterance has no history ($\nhisdecode [1] = 0$), second has $\nhisdecode [2] = 1$, and the 10th utterance has $\nhisdecode [10] = 3$.}
Although we can achieve much longer history, it is a huge burden for training as the history length grows and meanwhile causes serialization data training rather than the random shuffling one.

\section{SLongFNT: Speed up LongFNT ASR in streaming scenario}
\label{sec:streaming_longform}

In streaming ASR, besides the recognition error rate, the recognition latency stands as a pivotal metric.
{However, the LongFNT approach yields high latency, which inspire us to explore a streaming model that makes efficient use of historical information and keep the benfits of streaming models.}

{
Firstly, we introduce a modified version of the FNT architecture for streaming scenarios named as streaming FNT~(SFNT).
For speech encoder, we adopt the streaming conformer~\cite{gulati2020conformer} derived from its offline version.
Different from emformer~\cite{shi2021emformer}, our attention mechanism omits the memory bank and right-context (named as chunk-based attention~\cite{moritz21_interspeech}).
During training, the left-context and center-context are concatenated together as a large chunk.
During decoding, the left-context and its related states are consistently cached so that the chunked attention is like:
\begin{align}
    \bH_t = \text{MHA} (\bH_t,\bH_{\{u:v\}}, \bH_{\{u:v\}}),
\end{align}
where $t \in [u:v]$ {denotes the frames used in computation which starts from timestamp $u$ and ends at $v$, and $v-u$ is the summed length of left-context frames and center-context frames}.
Additionally, we substitute the traditional convolution block with a causal one.
}

{
As for $\predv$, as the use of transformer as $\predv$ can cause unacceptable delays, so we use LSTM structure~\cite{Hochreiter1997LongSM,chen2021developing} to improve the latency.
}

\subsection{SLongFNT-Text}
\label{sec:slongfnt_text}

{
As mentioned in Section~\ref{sec:longfnt_text}, the context encoder however is not efficient enough in terms of real time latency when we apply transformer-style architecture such as RoBERTa.
Thus, we propose an alternative approach: taking $\predv$ as the context encoder, rather than the transformer-style context encoder referenced in LongFNT~(transformer or RoBERTa).
This allows us to mitigate the computational delays associated with using an extra context encoder to process historical utterances.
}

\begin{figure}[ht]
  \centering
  \includegraphics[width=0.75\linewidth]{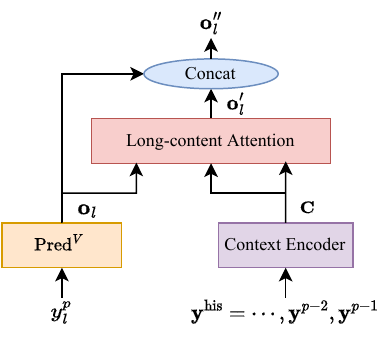}
  \caption{{Architecture of SLongFNT-Text: the long-content self-attention module and two different kinds of historical textual information.}}
  \label{fig:slongfnt-text}
\end{figure}

Shown in Figure~\ref{fig:slongfnt-text}(a), we improve the token-level integration mentioned in Section~\ref{sec:longfnt_text} from concatenating operation to the long-content attention:
\begin{align}
    \bo_l' = \text{Long-content Attention} (\bo_l, \bC, \bC),
\end{align}
where $\bo_l$ is the output state of $\predv$ before the projection layer, i.e. $\bz_l^V = \text{log\_softmax}(\text{Linear}^V (\bo_l))$, and $\bC$ is the context embedding sequence, {which is referred as RoBERTa or $\predv$ in experiments.}
{When using $\predv$ as the context encoder, and the hidden states of $\predv$ in SFNT is regraded as token-level embeddings $\bC$.}
Moreover, the {historical-enhanced} representation $\bo_l'$ is concatenated with $\bo_l$, $\bo_l'' = [\bo_l, \bo_l']$ to achieve better integration performance.

\subsection{SLongFNT-Speech}
\label{sec:slongfnt_speech}

\begin{figure}[ht]
  \centering
  \includegraphics[width=\linewidth]{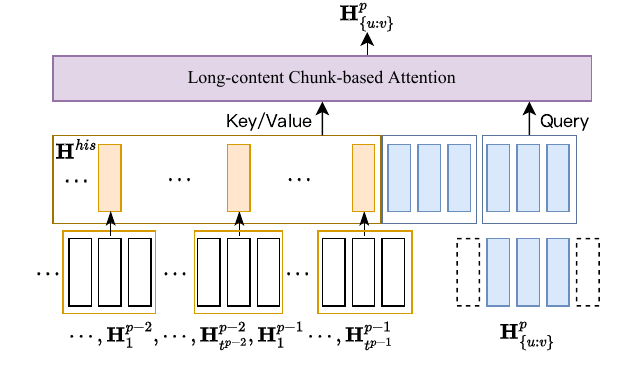}
  \caption{Architecture of {the attention layer} in SLongFNT-Speech: the {long-content} chunk-based self-attention module in the {speech encoder}. Downsampling is applied to reduce the history length.}
  \label{fig:slongfnt-speech}
\end{figure}

In the streaming FNT mentioned above, the encoder is used with limited acoustic frames in order to ensure low recognition latency, which also leads to a relatively large reduction in recognition accuracy.
Therefore, how to use historical information efficiently is explored in a streaming situation and a \textit{SLongFNT-Speech} architecture is proposed.
Shown in Figure~\ref{fig:slongfnt-speech}, when feeding external historical speech into the encoder, a similar technique called \textit{long-content chunk-based attention} can be applied to retain its historical features:
\begin{align}
    {\bH_t = \text{MHA} (\bH_t, [\bH^{\text{his}}; \bH_{\{u:v\}}],  [\bH^{\text{his}}; \bH_{\{u:v\}}])}.
    \label{eq:sfnt_speech}
\end{align}
where ${\bH^{\text{his}}}$ is the cached features from {historical utterances} $\cdots, {\bH^{p-2}}, {\bH^{p-1}}$. 
This allows the chunk-based attention to be expanded to a larger global field of perception.

Even though caching technique is applied in {long-content chunk-based} attention, we notice that the computation complexity $O(d \times (v-u + \cdots + t^{p-2} + t^{p-1}))$ is still huge, where the historical feature lengths $t^{p-2}, t^{p-1}$ is much longer than the chunk size.
This will result in a corresponding increase in computation when performing the attention calculation.
Accordingly two downsampling methods are proposed to reduce the increment length for historical acoustic information by rate $K$ into an acceptable magnitude shown in Figure~\ref{fig:slongfnt-speech}.

\textbf{Statistical Downsampling}:
The first method is called statistical downsampling. The uncompressed features $\bh^{<p}$ are firstly broken into discontinuous blocks, and then all frame features in each block are added and normalized to obtain a global representation, i.e. {block-wise mean or standard variance}.
{If the whole history is regraded as one block, then the downsampling can be seen as a global mean or standard variance.} For each block:
\begin{align}
    {\tilde{\bH}_i} = 1 / K \sum_{{K \cdot i \le t < K \cdot (i+1)}} {\bH_t},
\end{align}
where we compute the $i$-th block historical embedding, and $K$ represents the number of blocks from historical content.
Then $\bH^{his} = [\tilde{\bH}_1; \cdots ;\tilde{\bH}_K]$ mentioned in Equation~\ref{eq:sfnt_speech}.

\textbf{Dilated Downsampling}: Another method is called dilated local downsampling, where it is implemented by random selection:
\begin{align}
    {\tilde{\bH}_i} = {\bH_t} \text{, where } t \overset{uniform}{\leftarrow} {[K \cdot i, K \cdot (i+1))}.
\end{align}

By utilizing these two downsampling methods in training, and the statistical downsampling method in inference (which is referred to as `mix' in the experiments section), the proposed SLongFNT-Speech model can achieve better results.

\subsection{Training strategies for SLongFNT}

Combining SLongFNT-Text and SLongFNT-Speech, the final proposed streaming version of LongFNT is named \textbf{SLongFNT}~(Streaming LongFNT).
Firstly, we extend the previous FNT model in Section~\ref{sec:fnt} into the streaming version.
We mainly follow the setup from \cite{chen2021developing}, using truncated history with no future information.
Meanwhile, we follow MoCHA~\cite{chiu2018monotonic} to segment speech into chunks with a specified chunk size and attention caching used to optimize inference speed.
In addition, for the convolutional layer in the conformer, we also use causal convolution to ensure that future information will not be utilized in the training process.
The training is performed on concatenated utterances, and it enforces time restriction on the self-attention layers by masking attention weights, which can simulate a situation where future content is not available while still considering several look-ahead frames.

\section{Experimental Setup}
\label{sec:expr_setup}

We conduct experiments with two datasets, LibriSpeech~\cite{panayotov2015librispeech} and GigaSpeech Middle~(abbr. as GigaSpeech)~\cite{chen2021gigaspeech}.
LibriSpeech has around 960 hours of audiobook speech, while GigaSpeech has around 1,000 hours of audiobook, podcasting, and YouTube audio.
The sampling rate of these two datasets is 16 kHz.
The word error rate~(WER) averaged over each test set is reported.
For acoustic feature extraction, 80-dimensional mel filterbank~(Fbank) features are extracted with global level cepstral mean and variance normalization. Frame length and frame shift are 25ms and 10ms respectively.
Standard SpecAugment~\cite{park2019specaugment} is applied for both datasets.
Each utterance has two frequency masks with parameter ($F=27$) and ten time masks with maximum time-mask ratio ($pS = 0.05$).
5,000 word pieces with Byte Pair Encoding~(BPE)~\cite{kudo2018sentencepiece} are trained using LibriSpeech and GigaSpeech datasets separately.
As for the $\predv$ part, the text scale is 10.27 million words for LibriSpeech and 9.68 million for GigaSpeech.
And the extra text data (`+ external text' mentioned in Section \ref{sec:longfnt_text}), for LibriSpeech, we use the {official} extra text corpus which has 812.69 million words {(https://www.openslr.org/11/)}, and for GigaSpeech, we use GigaSpeech-XL training text data which has 113.80 million words.
{The external text is used to pre-train $\predv$ to get a better initialization.}

\begin{figure}[ht]
  \centering
  \includegraphics[width=\linewidth]{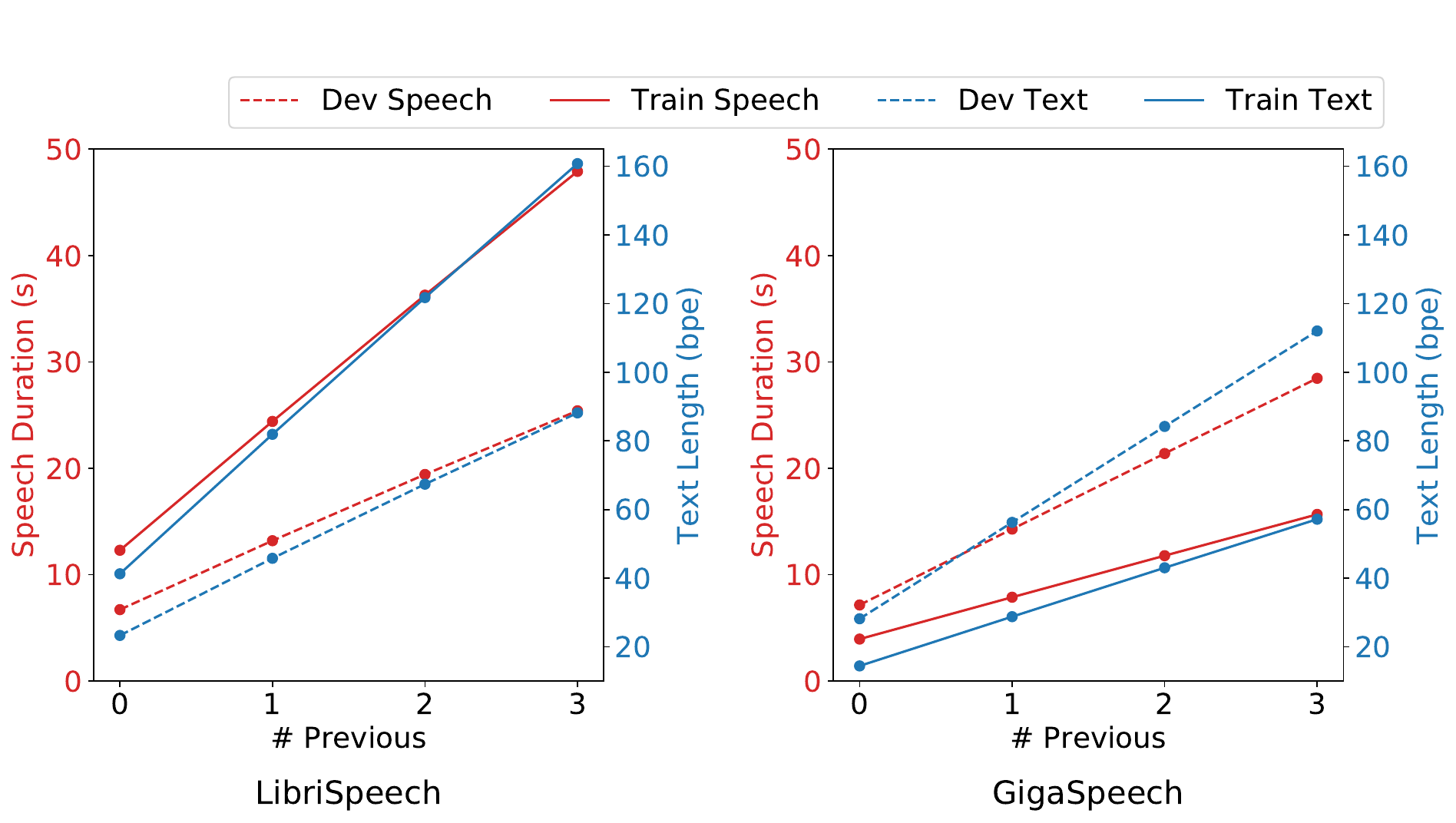}
  \caption{The statistics of speech and words in the LibriSpeech~(left) and GigaSpeech~(right). {\color{red} Red} denotes speech, and {\color{blue} blue} denotes text. Solid line denotes train set and dotted line denotes dev set. The left vertical axis of each figure denotes the speech duration (ms), while the right one denotes the text length (bpe5000). The horizontal axis \#Previous denotes $\nhis$.}
  \label{fig:dataset}
\end{figure}

 
As shown in Figure~\ref{fig:dataset}, we evaluate the average length of audio and bpe-level tokens in different {long-content} setups, i.e. the different numbers of historical utterances.
{
In our experiments utilizing Librispeech and Gigaspeech, we maintained the temporal sequence of sentences using successive utterance ids, such as XXX\_01, XXX\_02 and etc.
Gigaspeech occasionally exhibits sequence discontinuities, but given their typical session lengths over one hour, such breaks have a marginal impact on continuity.
While not every Librispeech book's content is fully represented, each session's content is intact and sequential.
During testing, for any utterance with non-consecutive historical utterances, we adaptively treat the immediately preceding utterance as historical content to balance both relevance and efficiency.
Take `03,05,06' utterance sequence as an example, during the inference of '06' with $\nhis=2$ and '04' is missing, we would consider '03' and '05' as the preceding historical content for '06'.
For the first utterance, there is no historical information, whereas the second utterance has only '01' as its historical content.
}

The non-streaming {FNT} baseline follows the {single-utterance} settings of factorized neural transducer (FNT) \cite{zhaorui}.
{And the non-streaming C-T baseline has the same architecture of speech encoder and predictor and training setup where as FNT, the predictor has the same shape as $\predb$.}
The subsampling layer is a VGG2L-like network, which contains four convolution layers with the down-sampling rate of 4.
The encoder has 18 conformer layers, in which the inner size of the feed-forward layer is 1,024, and the attention dimension is 512 with 8 heads.
The $\predb$ has two unidirectional LSTM layers with 1,024 hidden size and the joint dimension is set to 512.
The $\predv$ use vanilla 8 transformer layers, which has 256 attention dimension with 8 heads.
{The textual input for $\predb$ and $\predv$ is prefixed with a start-of-sequence~(SOS) token.}
The hyper-parameter weights are fixed as $\lambda_{CTC}=0.1$.
The context encoder has the same shape as $\predv$ if training from scratch, and utilizes the frozen RoBERTa model otherwise.
{The input of the context encoder is also always started with a start-of-sequence~(SOS) symbol for distinction.}
During inference, we keep beam size equal to 8.
As for the FNT's external larger text-trained LM~(external text), we use 16 transformer layers with 512 attention dimension with 8 heads.
{In the following experiments, the default number of long-content sentences $\nhis$ is two. We also conducted ablation studies to explore the effects of varying this number.}
{Throughout both training and decoding, we consistently incorporate $\le \nhis$ sentences as the historical textual content.}
Two possible {long-content} text forms, i.e. oracle and hypotheses, are fed into LongFNT model {series} to obtain the results.
{
Both FNT and LongFNT follow consistent training stages.
When no external text is utilized, both models are trained from scratch.
However, when `+external text' is incorporated, the $\predv$ is pre-trained using the external text data.
}

As for the streaming experiments, the basic training/inference setup remains the same, but the conformer encoder is with chunk-wise casual convolution layer.
We train the streaming encoder with {8 left chunks and 1 center chunk} (per chunk is 40ms, i.e. totally 320ms delay in the basic streaming FNT system). 
Furthermore, in streaming human-computer interaction scenarios, we have to take the speech time into consideration when evaluating the real decoding time.
Under this scenario, the non-streaming ASR system will be suspended until all the audio frames have been received, which will cause high latency, while the streaming one can process the speech as far as it receives the audio chunk.
{And the training stages for SFNT/SLongFNT remains the same as FNT/LongFNT for `+ long text', `+ external text' and etc.}

We measure the end-latency under single core of Intel(R) Xeon(R) Gold 6132 CPU @ 2.60GHz:
\begin{align}
    {\text{End-Latency}^p = T^{p}_{\text{end}} - T^{p}},
\end{align}
{
where $T^p$ is the length of utterance $p$, and $T^p_{\text{end}}$ denotes the total computation time elapsed from the moment the first frame is fed to the encoder until the final bpe token is decoded for utterance $p$.
Meanwhile, we evaluate the average end-latency for the whole dev set.
}
To be notified, the context embedding sequence $\bC$ is simulated to compute in another CPU core to avoid any latency issues during the inference process.

\section{Experimental Results and Analysis}
\label{sec:expr}

\subsection{Evaluation on the non-streaming LongFNT model}
\label{sec:expr_longfnt}

In our initial experiments, we investigated whether the performance of the vanilla C-T model could be improved by incorporating {long-content} transcription history~(`+ long text'), in the first block of Table~\ref{tab:longfnt}.
We observed that the vanilla C-T model demonstrated only minimal improvements when using extended {long-content} text or utterance-level integration, even with the use of oracle transcripts.
Moreover, when incorporating hypotheses, the system performance actually degraded.
This outcome can be attributed to the fact that the predictor network in traditional neural transducer models is not a pure language model and cannot easily leverage longer history to achieve performance improvements.

\begin{table}[ht]
  \centering
  \caption{Performance~(WER)~(\%) comparison on LibriSpeech test sets and GigaSpeech dev/test sets under the non-streaming scenario.
  Text has two modes, `oracle' denotes that the historical text is obtained from the ground truth transcripts, while `hypotheses' denotes that the historical text is from the decoding results.
  } 
  \label{tab:longfnt}
  \begin{tabular}{l|l|cccc}
    \toprule
    \multirow{2}{*}{Model} & \multirow{2}{*}{Text} & \multicolumn{2}{c}{Libri-test} & \multicolumn{2}{c}{Giga} \\
    && clean & other & dev & test \\
    \midrule
    \midrule
    C-T & - & 3.1 & 6.6 & 16.1 & 15.7 \\
    + long text & oracle & 3.1 & 6.5 & 15.8 & 15.5 \\
    + long text & hyp & 3.1 & 6.7 & 16.1 & 15.9 \\
    + utterance-level integ. & oracle & 3.1 & 6.5 & 16.0 & 15.5 \\
    + utterance-level integ. & hyp & 3.2 & 6.9 & 16.4 & 16.2 \\
    \midrule
    FNT       & - & 3.2 & 6.4 & 16.8 & 16.3 \\
    + long text & oracle  & 3.2 & 6.3 & 16.5 & 16.2 \\
    + long text & hyp & 3.2 & 6.4 & 16.7 & 16.4 \\
    + external text   & -    & 3.0 & 6.1 & 16.4 & 16.0 \\
    + external + long text & hyp     & 3.0 & 6.1 & 16.4 & 16.1 \\
    {+ long speech} & - & {3.2} & {6.5} & {17.0} & {16.6} \\
    \midrule
    \multicolumn{6}{l}{FNT \textit{with LM fusion}} \\
    {+ Single-utterance LM} & {-} & {2.6} & {5.7} & {16.0} & {15.6} \\
    {+ Cross-utterance LM \cite{irie2019trainingb,chen2020lstm}} & {oracle} & {2.2} & {5.1} & {14.6} & {14.3} \\
    {+ Cross-utterance LM \cite{irie2019trainingb,chen2020lstm}} & {hyp} & {2.4} & {5.5} & {15.5} & {15.1} \\
    \midrule
    {LongFNT-Text} & {hyp} & {2.5} & {5.5} & {15.2} & {14.9} \\
    {+ LM Fusion} & {hyp} & {2.2} & {4.8} & {14.4} & {14.0} \\
    LongFNT-Speech & -        & 2.8 & 6.0 & 15.9 & 15.7 \\
    LongFNT  & hyp     & \textbf{2.4} & \textbf{5.4} & \textbf{14.8} & \textbf{14.3} \\
    \bottomrule
  \end{tabular}
\end{table}

In terms of FNT, shown in the second block of Table~\ref{tab:longfnt}, including the decoded/oracle {long-content} transcription history in the text input~(`+ long text') does very limited help on both LibriSpeech and GigaSpeech. When the factorized predictor V is trained with extra text data~(+external text), the system can obtain small but consistent improvements on all test sets. It demonstrates that the FNT architecture can factorize and model the language knowledge more accurately and can be benefited from a powerful language model. However, the {long-content} text history also does not further improve system upon the external text, which indicates that it is non-trivial to explore how to better leverage {long-content} information in the neural transducer. 
{Additionally, while the normal speech encoder is indeed capable of handling long-form speech, our experiment `FNT + long speech' reveals that directly integrating long-content speech with the FNT results in a performance drop across all sets.
This underscore the need to find a more optimal approach for leveraging long-content speech.}

In the last, the newly proposed LongFNT model is evaluated and the results are shown in the last block of Table~\ref{tab:longfnt}. Compared to the previous C-T and FNT systems, all the proposed models using long information can get significant improvements, which demonstrates the better utilization of {long-content} history information with the new model.
{In contrast to `FNT + long speech', LongFNT-Speech exhibits superior performance.
This enhancement can be attributed to our strategy of confining the gradient back-propagation.}
{To further validate the efficacy of LongFNT-Text, we replicated the cross-utterance transformer LM as detailed in \cite{zeyer2020new,chen2020lstm} in the third block of Table~\ref{tab:longfnt}.
Nonetheless, it is worth noting that these methods aren't entirely analogous to LongFNT-Text.
The primary distinction arises from their reliance on shallow fusion, whereas LongFNT-Text operates independently of such a fusion mechanism.
Shown in Table~\ref{tab:longfnt}, it is evident that cross-utterance LM fusion (hyp) garners an improvement over conventional fusion and baseline.
Meanwhile, LongFNT-Text outperforms cross-utterance LM fusion on GigaSpeech, possibly due to its enhanced robustness.
Furthermore, when LongFNT-Text is applied with shallow fusion, its performance notably surpasses that of `+cross-utterance LM fusion (hyp)'.
Such a synergy effectively captures the long-content textual information.}
Furthermore, it is observed that both two types of historical information, i.e. text and speech, are both helpful for {long-content} speech recognition, and LongFNT-Text is slightly better than the LongFNT-Speech. The final LongFNT model integrating historical knowledge from both text and speech achieves the best system performance, and compared to the limited gain from the naive long text usage in C-T and FNT, the proposed LongFNT obtains around 20\% and 10\% relative WER reductions on LibriSpeech and GigaSpeech respectively.

\subsection{Ablation Studies on the non-streaming LongFNT model}
\label{sec:expr_longfnt_abl}

\subsubsection{The Number of Historical {Utterances}}
\label{sec:expr_longfnt_num}

Performance comparison of successive historical utterance counts are evaluated with the proposed LongFNT model, and the results are shown in Table~\ref{tab:longfnt_num}. The correlation between the number of historical utterances and the recognition error rate can be observed, and $\nhis$=0 means no historical utterance is used, i.e. the FNT baseline.
As $\nhis$ grows, the WER of the current utterance is reduced gradually with the increased historical length $\nhis$.
After $\text{$\nhis$} > 2$, the improvement is limited but training resources are very consumed. So the previous two utterances history are the most appropriate trade-off point between accuracy and cost in LongFNT, and we set $\nhis$=2 to learn appropriate {long-content} information for all the following experiments.

\begin{table}[ht]
  \centering
  \caption{Performance~(WER)~(\%) comparison of successive historical utterance counts on GigaSpeech dev/test sets for LongFNT series.
  {Specially, $\nhis$ for the proposed LongFNT model series is decoded under $\nhistrain = 2$ while $\nhisdecode = 0$ as in Section~\ref{sec:longfnt_train}.}
  All results are evaluated using decoded hypotheses as the historical text.
  }
  \label{tab:longfnt_num}
  \begin{tabular}{l|cccc}
    \toprule
    $\nhis$ & 0 & 1 & 2 & 3 \\
    \midrule
    FNT            & 16.8/16.3 & - & - & - \\
    LongFNT-Text   & {17.0/16.4} & 16.0/15.6 & 15.2/14.9 & 15.3/14.8 \\
    LongFNT-Speech & {16.7/16.3} & 16.2/15.8 & 15.9/15.7 & 15.8/15.5 \\
    LongFNT        & {16.9/16.4} & 15.8/15.4 & \textit{14.8/14.3} & 14.7/14.3 \\
    \bottomrule
  \end{tabular}
\end{table}

\subsubsection{Effectiveness of different components in LongFNT-Text}

\begin{table}[ht]
  \centering
  \caption{Performance~(WER)~(\%) comparison of different components in LongFNT-Text (The final LongFNT-Text system is the line denoted with *).}
  \label{tab:longfnt_text}
  \begin{tabular}{l|ccccc}
    \toprule
    \multirow{2}{*}{Model} & \multirow{2}{*}{Text} & \multicolumn{2}{c}{Libri-test} & \multicolumn{2}{c}{Giga} \\
    && clean & other & dev & test \\
    \toprule
    FNT &-& 3.2 & 6.4 & 16.8 & 16.3 \\
    \midrule
    \texttt{+} utterance-level integ. & oracle & 2.9 & 6.1 & 16.0 & 15.8 \\ 
    \texttt{+} utterance-level integ. & hyp & 3.0 & 6.3 & 16.5 & 16.2 \\ 
    \texttt{+} token-level integ.    & oracle & 2.9 & 6.0 & 15.7 & 15.4 \\
    \texttt{+} token-level integ.    & hyp & 3.0 & 6.1 & 16.0 & 15.7 \\
    \midrule
    \texttt{+} external text & - & 3.0 & 6.1 & 16.4 & 16.0 \\
    \texttt{++} utterance-level integ. & hyp & 2.9 & 5.9 & 16.0 & 15.5 \\
    \texttt{+++} RoBERTa & hyp & 2.8 & 5.8 & 15.9 & 15.3 \\
    \texttt{++} token-level integ. & hyp & 2.6 & 5.7 & 15.8 & 15.4 \\
    \texttt{+++} RoBERTa & hyp & 2.6 & 5.6 & 15.8 & 15.4 \\
    \midrule
    \texttt{+} utterance- \texttt{+} token- integ. & hyp & 2.8 & 5.8 & 15.9 & 15.5 \\
    \texttt{++} external text & hyp & 2.5 & 5.6 & 15.6 & 15.2 \\
    \texttt{+++} RoBERTa (*) & hyp & 2.5 & 5.5 & 15.2 & 14.9 \\
    \bottomrule
  \end{tabular}
\end{table}

In this subsection, we evaluate the effectiveness of the proposed LongFNT model and explore the impact of each module on the final performance.
Table~\ref{tab:longfnt_text} presents the results of our analysis on the efficiency of different integration methods for LongFNT-Text.

In the first block of Table~\ref{tab:longfnt_text}, experiments indicate that token-level integration is more significant than utterance-level integration.
When considering only utterance-level integration, the system achieves a WER reduction of 0.2/0.1 absolute on the LibriSpeech and 0.3/0.1 absolute on the GigaSpeech datasets.
However, when token-level integration is utilized, the WER reduction improves to 0.2/0.3 absolute on the LibriSpeech corpus and 0.8/0.6 absolute on the GigaSpeech corpus.
A similar trend can be observed in the second block, token-level integration outperforms utterance-level integration by $\sim$0.4 absolute WER reduction on LibriSpeech and 0.2$\sim$0.5 on GigaSpeech, after adding external text~(i.e. with large LM, `+ external text') and after using RoBERTa model~(`\texttt{+++} RoBERTa').

As mentioned previously in Section~\ref{sec:expr_setup}, in the real scenario, ground truth~(oracle) text can not be accessed, and we evaluate the above methods using decoded transcriptions~(hypotheses) to get real performance and explore the importance of those two types in different LongFNT-Text modes.
Shown as the 1st block of Table~\ref{tab:longfnt_text}, experiments indicate that models decoded using hypotheses experienced a relative 1\%$\sim$5\% performance drop compared to those decoded using oracle text, and the degradation on the GigaSpeech is larger compared to the that on the LibriSpeech.
This may be attributed to the fact that the basic error rate influences the performance of hypotheses, and a system with low WER is necessary to achieve performance improvements in {long-content} speech recognition.
Additionally, we observe that the utterance-level integration is more sensitive to the {long-content} transcriptions quality compared to the token-level one, as it drops more 0.1\%$\sim$0.2\% absolute WER.

This indicates the shortcoming of statistical averaging pooling for utterance-level integration.

Then we discovered that {long-content} transcriptions can be effectively utilized in conjunction with external text, thereby leveraging the advantages of FNT.
Results in the 2nd and 3rd blocks of Table~\ref{tab:longfnt_text} demonstrate a relative performance increase of at least 2\% across all datasets for different textual integration methods (`+token-level', `+ utterance-level', `+ utterance-level + token-level').
These results highlight the effectiveness of employing a external-text-boosted vocabulary predictor to further enhance the performance of our models.

Furthermore, We also investigated the importance of replacing the train-from-scratch context encoder with a pre-trained RoBERTa model, and found that the importance varied across the LibriSpeech and GigaSpeech datasets.
For LibriSpeech, the performance only improves by 0.1\% absolute WER reduction, and in some cases, no improvement was observed in the test-clean set for the LongFNT-Text model.
In contrast, for GigaSpeech, we observed a consistent improvement in performance of at least 0.3\%$\sim$0.5\% absolute WER reduction.
This phenomenon is interesting because the impact of the RoBERTa model was minimal in the LibriSpeech dataset.
This may be attributed to the fact that the influence of transcriptions is relatively smaller in this dataset, and the context encoder has a similar ability to model the {long-content} textual information.

\subsection{Evaluation on streaming SLongFNT model}
\label{sec:expr_slongfnt}

Table~\ref{tab:slongfnt} presents the performance comparison in streaming scenario, and here we no longer explore the performance of C-T with the long-term history (the specific reasons can be seen in Section~\ref{sec:expr_longfnt}).
We first set a benchmark based on the impact of {long-content} text on the original FNT model, which the basic streaming FNT (SFNT) has obvious performance drop compared with the non-streaming FNT model in Table~\ref{tab:longfnt}.
The results presented in lines 1-3 of the first block of the Table~\ref{tab:slongfnt} align with the conclusions drawn in Section~\ref{sec:expr_longfnt} for non-streaming system with oracle/hypotheses historical condition.
With the help of external text, whether there is {long-content} text or not, the streaming FNT model has been improved consistently, although the improvement is relatively small.

\begin{table}[ht]
  \centering
  \caption{performance (WER) (\%) comparison on LibriSpeech test sets and GigaSpeech dev/test sets under the streaming scenario. Text has two modes, `oracle' denotes that the historical text is obtained from ground truth transcripts, while `hyp' denotes that the historical text is from the decoded hypotheses.}
  \label{tab:slongfnt}
  \begin{tabular}{l|l|cccc}
    \toprule
    \multirow{2}{*}{Model} & \multirow{2}{*}{Text} & \multicolumn{2}{c}{Libri-test} & \multicolumn{2}{c}{Giga} \\
    && clean & other & dev & test \\
    \midrule
    SFNT                    &-         & 4.1 & 10.0 & 22.9 & 22.0 \\
    + long text            &oracle    & 3.8 & 9.3  & 21.4 & 20.6 \\
    + long text            &hyp & 4.0 & 9.8  & 22.8 & 22.2 \\
    + external text        &- & 3.9 & 9.7  & 22.3 & 21.5 \\
    + external + long text &hyp& 3.8 & 9.4  & 21.9 & 21.0 \\

    \midrule
    SLongFNT-Text   & hyp & 3.7 & 8.8 & 21.2 & 20.0 \\
    SLongFNT-Speech & -          & 3.3 & 7.8 & 19.7 & 18.8 \\
    SLongFNT        & hyp & \textbf{3.1} & \textbf{7.5} & \textbf{19.1} & \textbf{18.2} \\
    \bottomrule
  \end{tabular}
\end{table}

In the second block of Table~\ref{tab:slongfnt}, it shows the performance of our proposed SLongFNT models individually.
For the utilization of {long-content} transcriptions, i.e. SLongFNT-Text, it achieves 11/8\% relative WER reduction on LibriSpeech/GigaSpeech, which is a comparable improvement with non-streaming LongFNT-Text.
As for the utilization of {long-content} speech, i.e. SLongFNT-Speech, the downsampling rate is set to $K=4$ to consider the balance between the inference speed and accuracy, and 15/14\% relative WER reduction on LibriSpeech/GigaSpeech are observed.
It is found that the improvement of SLongFNT-Speech is larger than that of LongFNT-Speech, which may be due to the greater importance of speech encoder in the streaming condition.
Finally, the SLongFNT system combining historical knowledge from both text and speech achieves $\sim$25\% relative WERR on LibriSpeech and $\sim$17\% relative WERR on GigaSpeech, compared to the baseline streaming FNT system.

\subsection{Ablation Studies on the streaming SLongFNT model}

At first, we follow the $\nhis=2$ setup for the exploration of SLongFNT-Text/-Speech architecture to find optimal architecture setup for SLongFNT.

\begin{table}[ht]
  \centering
  \caption{The ablation study of different components proposed in SLongFNT. For SLongFNT-Text, the second block shows different integration methods. For SLongFNT-Speech, the third block shows different down-sampling strategies.}
  \label{tab:slongfnt_ablation}
  \begin{tabular}{l|ccccc}
    \toprule
    \multirow{2}{*}{Model} & \multicolumn{2}{c}{Libri-test} & \multicolumn{2}{c}{Giga}  \\
    & clean & other & dev & test \\
    \midrule
    SFNT           & 4.1 & 10.0 & 22.9 & 22.0 \\
    SLongFNT       & 3.1 & 7.5  & 19.1 & 18.2 \\
    
    \midrule
    \multicolumn{5}{l}{\textit{SLongFNT-Text (w/ RoBERTa or $\predv$ context encoder})} \\
    w/  RoBERTa (oracle)         & 3.7 & 8.9 & 19.2 & 18.3 \\
    w/  RoBERTa (hyp)     & 3.9 & 9.5 & 21.6 & 20.8 \\
    + external text     & 3.7 & 8.8 & 21.2 & 20.0 \\

    w/  $\predv$ (oracle)        & 3.8 & 9.1 & 19.5 & 18.4 \\
    w/  $\predv$ (hyp)    & 4.0 & 9.8 & 22.1 & 21.2 \\
    + external text     & 3.9 & 9.6 & 21.8 & 20.9 \\

    \midrule
    \multicolumn{5}{l}{\textit{SLongFNT-Speech~(w/ downsampling methods)}} \\
    w/o downsample (K=1) & 3.2 & 7.5 & 19.4 & 18.6 \\
    mean                 & 4.0 & 9.6 & 22.3 & 21.5 \\
    mean+std             & 4.0 & 9.8 & 22.5 & 21.6 \\
    statistical (K=2)    & 3.3 & 7.6 & 19.5 & 18.6 \\
    statistical (K=4)    & 3.3 & 7.9 & 19.9 & 19.1 \\
    statistical (K=8)    & 3.5 & 8.2 & 20.7 & 20.3 \\
    statistical (K=16)   & 3.8 & 9.3 & 22.0 & 21.2 \\ 
    dilated (K=2)        & 3.3 & 7.9 & 19.9 & 19.1 \\
    dilated (K=4)        & 3.4 & 8.0 & 20.2 & 19.4 \\
    mix (K=4)            & \textbf{3.3} & \textbf{7.8} & \textbf{19.7} & \textbf{18.8} \\
    \bottomrule
  \end{tabular}
\end{table}

\subsubsection{Exploring different text integration methods for SLongFNT-Text}
\label{sec:expr_slongfnt_text}

In the second block of Table~\ref{tab:slongfnt_ablation}, we investigate different components of SLongFNT-Text.
First of all, we explore the external context encoder method~(directly use the same RoBERTa model mentioned in Section~\ref{sec:longfnt_text}), and the method of extracting context embedding is the same as LongFNT.
RoBERTa~(hypotheses) is consistently worse than RoBERTa~(oracle), and the drop is larger for datasets with worse WER (i.e. 7\% relative WER degradation on LibriSpeech while 14\% for GigaSpeech) due to the incorrect {textual content}.
Similar to LongFNT-Text, with the help of external text, the RoBERTa-based SLongFNT-Text model achieves $>$8\% relative WER improvement on both datasets.

However, in real streaming scenario, the CPU is overloaded for current chunk's inference, thus computing context embeddings by RoBERTa causes higher latency.
We then try to utilize the hidden state from $\predv$, which does not require external computing resources.
With $\predv$ as the internal context encoder, the final SLongFNT-Text also achieves significant improvement, and it is smaller than the RoBERTa context encoder but still obvious. 

\subsubsection{Exploring different downsampling methods for SLongFNT-Speech}
\label{sec:expr_slongfnt_speech}

In the third block of Table~\ref{tab:slongfnt_ablation}, we investigate different downsampling methods for SLongFNT-Speech mentioned in Section~\ref{sec:slongfnt_speech}, and downsampling rate $K=1$ indicates the system without downsampling.
We first considered the corner cases, those are, to calculate the mean vector of all {long-content} acoustic representations~(mean), and to calculate the mean and standard variance of all {long-content} acoustic representations~(mean+std).
Although these modes are simple, experimental results show that both methods can be still better than the SFNT baseline~(line 1), and the `mean' performs slightly better than the model with `mean+std'.
Thus we choose `mean' as the implementation of statistical downsampling.

The statistical and dilated downsampling with different rates are then applied. It is observed that the statistical downsampling with $K=2$ gets almost no WER degradation, and the performance drop will become obvious when $K>4$. Compared to the dilated downsampling mode, the statistical mode is consistently better.
Moreover, we tried to combine statistical and dilated downsampling modes with $K=4$, i.e. using the statistical with dilated downsampling methods during training and using statistical downsampling only for decoding, SLongFNT-Speech (shown as the last line in Table~\ref{tab:slongfnt_ablation}) can achieve better results and perform the appropriate trade-off point between the accuracy and computation cost.


\subsubsection{The Number of Historical {Utterances}}



We study the influence of different historical information lengths on the recognition accuracy and delay of the SLongFNT model, and then select the most suitable number for the following experiments.
From Table~\ref{tab:num_slongfnt},  it is found that as the number of historical utterances $\nhis$ increases, the performance of SLongFNT-Text/-Speech and the final model SLongFNT are gradually improved. The impact from historical speech is larger than that from historical text.
Similar as the observation for non-streaming LongFNT, $\nhis=2$ also seems be the appropriate trade-off point between accuracy and computational cost for this streaming SLongFNT, and it is applied on SLongFNT in the further experiments.

\begin{table}[ht]
  \centering
  \caption{Performance~(WER)~(\%) comparison of successive utterance counts $\nhis$ on GigaSpeech dev/test sets {for SLongFNT series.}}
  \label{tab:num_slongfnt}
  \begin{tabular}{l|cccc}
    \toprule
    $\nhis$ & 0 & 1 & 2 & 3 \\
    \midrule
    SFNT             & 22.9/22.0 & -  & - & - \\
    SLongFNT-Text   & {23.2/22.4} & 21.8/20.8 & 21.2/20.0 & 21.4/20.2 \\
    SLongFNT-Speech & {22.7/21.7} & 20.3/19.9 & 19.7/18.8 & 19.5/18.5 \\
    SLongFNT        & {23.1/22.3} & 19.9/19.4 & \textit{19.1/18.2} & 18.7/18.0 \\
    \bottomrule
  \end{tabular}
\end{table}

\begin{figure}[ht]
    \centering
    \includegraphics[width=0.8\linewidth]{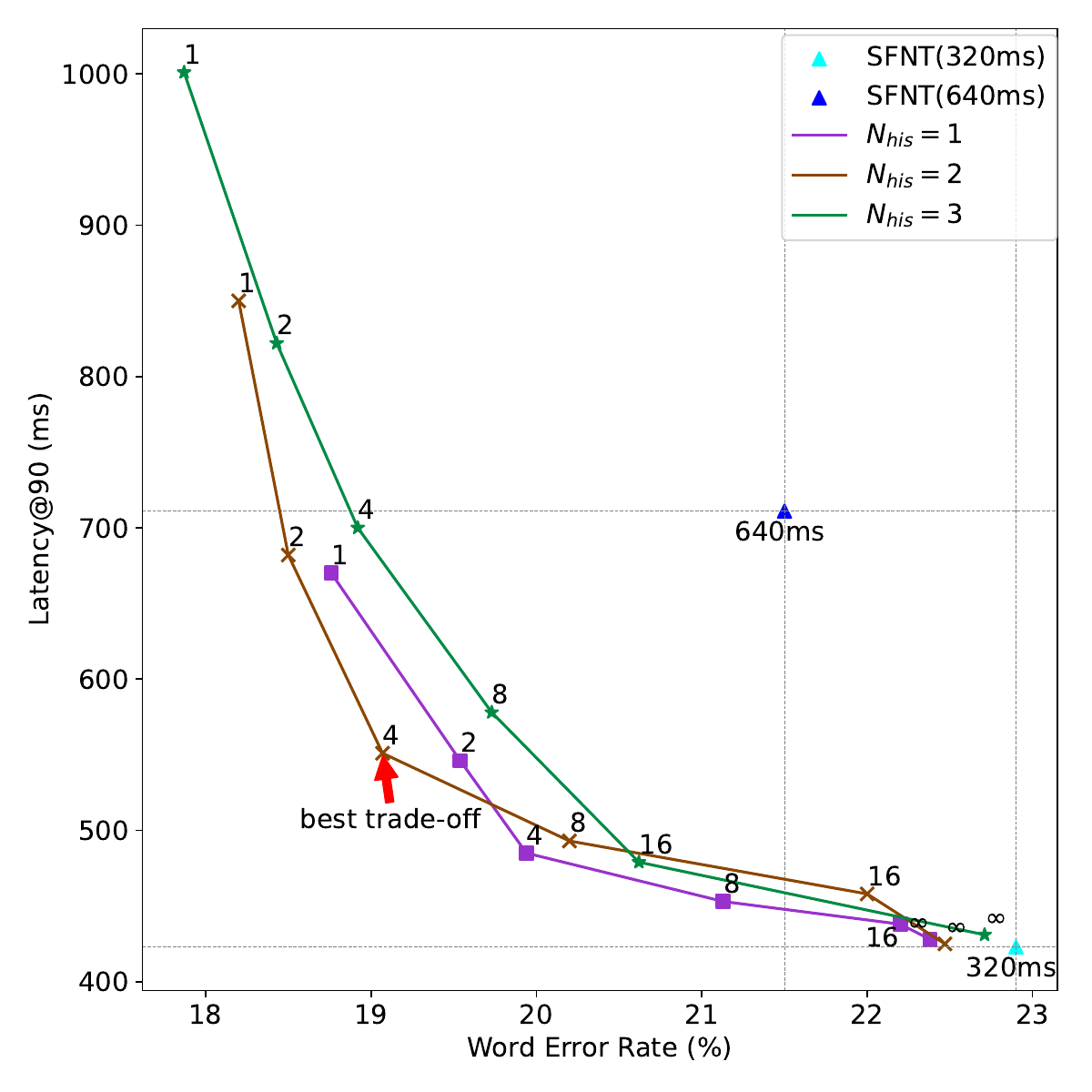}
    \caption{Latency~(ms) and word error rate~(\%) trade-off on dev set of GigaSpeech. $\infty$ denotes `mean' in Table~\ref{tab:slongfnt_ablation}. Different downsampling rates $K$ and $N_{his}$ are presented. The numbers on the line are the related downsampling rates $K$.}
    \label{fig:latency}
\end{figure}

\subsubsection{Decoding Latency}
Decoding speed is another concern when people deploy the streaming ASR systems.
Illustrated in Figure~\ref{fig:latency}, we evaluated the end-latency for the proposed SLongFNT systems with different number of historical utterances ($\nhis$ = 0,1,2,3) and different downsampling rates ($K$ = 1,2,4,8,16,$\infty$), where $\infty$ denotes mean pooling.
The results demonstrate that as the word error rate decreases, the latency increases.
We can balance the trade-off between latency and model accuracy (error rate) by choosing $\nhis=2, K=4$, which results in the latency of 545ms.
{In this setup, the latency for SLongFNT-Text is recorded at 473ms, while SLongFNT-Speech exhibits a latency of approximately 497ms.}

Meanwhile, the figure also demonstrates that the down-sampling ratio has a significant impact on latency, and the increase of historical statements gradually increases overall recognition delay.
When the downsampling ratio is greater than 4, the rate of latency attenuation gradually slows down.
At this point, the infinity downsampling latency is similar to $\nhis=0$, where we consider ‘mean’ as an approximation that approaches infinity ($\infty$).
Moreover, another basic streaming FNT model with larger chunk-size=640ms is also constructed and shown in the figure, denoted as SFNT(640ms). It is observed that the newly proposed SLongFNT with $\nhis=2 \& K = 2$, has the similar latency as SFNT~(640ms) but with much better accuracy.
It is worth noting that when taking the RoBERTa context encoder into consideration, an additional average latency of 200ms is added to the current model. This issue can be addressed by using $\predv$ as the context encoder, as mentioned in Section~\ref{sec:slongfnt_text}.

\subsection{Illustration on the {Long-Content} Speech Recognition Results}

\begin{figure*}[ht]
     \begin{subfigure}[b]{0.5\linewidth}
         \centering
         \includegraphics[width=\textwidth]{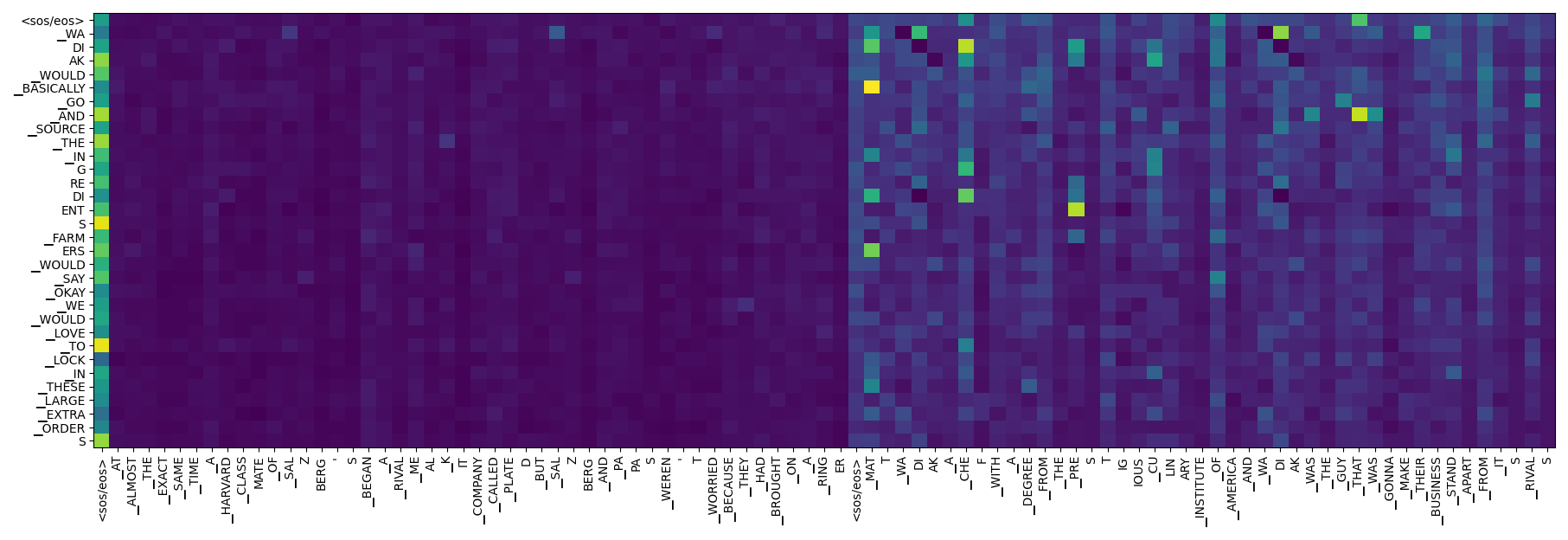}
         \caption{Bottom Layer}
         \label{fig:att_bottom}
     \end{subfigure}
     \hfill
     \begin{subfigure}[b]{0.5\linewidth}
         \centering
         \includegraphics[width=\textwidth]{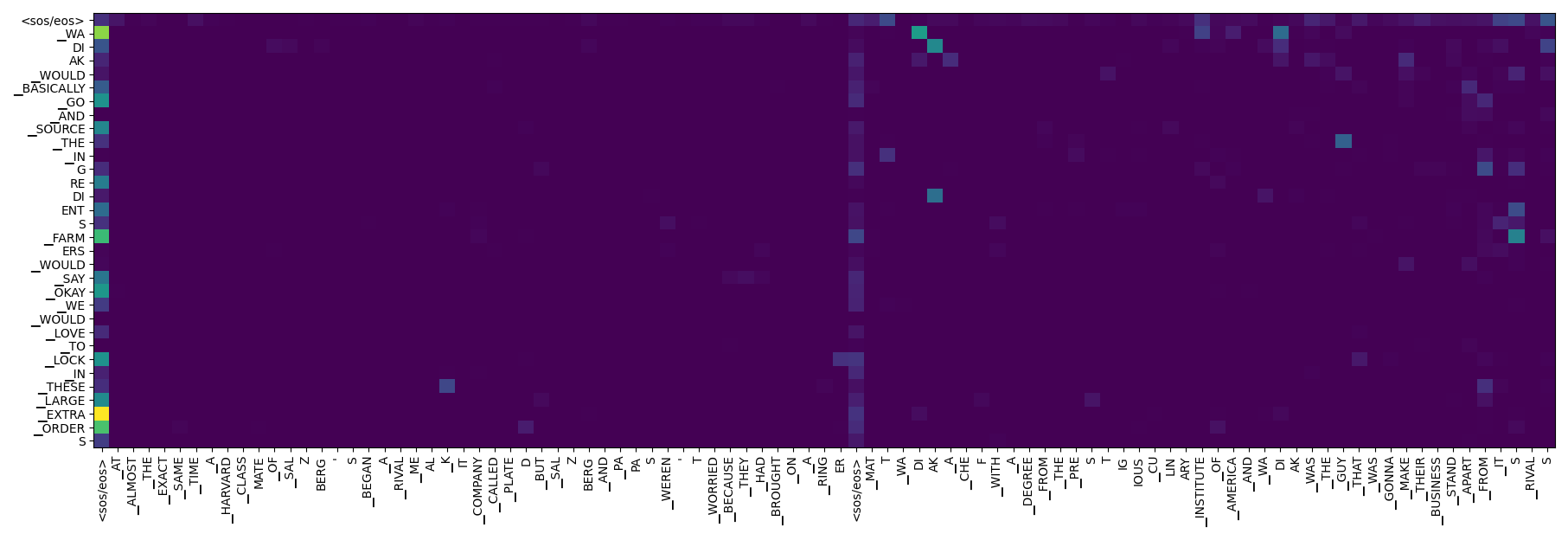}
         \caption{Top Layer}
         \label{fig:att_top}
     \end{subfigure}
     \caption{
     The {long-content} attention of token-level integration in $\predv$~($\nhis = 2$). The left is the attention in the bottom layer and the right is the attention in the top layer. The example is the same as the first block in Table~\ref{tab:egs}, where the utterance id is POD1000000004\_S0000075 in GigaSpeech {dev set}.
     {Its historical sentences are POD1000000004\_S0000073 and POD1000000004\_S0000074, with sentences separated by the SOS token ($<$sos/eos$>$).}
     }
     \label{fig:attention}
\end{figure*}

The Table~\ref{tab:egs} presents several examples that demonstrate how LongFNT corrects the results of normal FNT by utilizing {long-content} textual information.
For the first block, when using ground truth as the history, LongFNT successfully used the historical word `WADIAK', and makes the appropriate modification.
However, when using hypotheses as history (i.e. `WADIAC') that is incorrectly decoded, the LongFNT utilizes the wrong historical information and cannot correct the error `WADIAC'.
For the second and third blocks, when the hypotheses history are correct, LongFNT is also able to utilize the historical words and then make successful modifications.

\begin{table}[ht]
    \setlength{\tabcolsep}{3pt}
    
    \centering
    \caption{Examples of decoded results~(Hyp.) comparision between the normal FNT and proposed LongFNT model. For LongFNT, two types of text histories (Hist.), i.e. ground truth and hypotheses, are shown, and G.T. means the real ground truth transcription. {SOS token is ignored for simplification.}}
    \label{tab:egs}
    \begin{tabular}{@{}l|p{0.8\linewidth}@{}}
    \toprule
    Hist.gt & ... matt \textcolor{ogreen}{WADIAK} a chef with a degree from the prestigious culinary institute \\
    Hist.hyp& ... matt \textcolor{red}{WADIAC} a chef with a degree from the prestigious culinary institute \\
    G.T.& \textcolor{blue}{WADIAK} would basically go and source the ingredients ... \\
    Hyp.FNT& \textcolor{red}{WADIAC} would basically go and source the ingredients... \\
    Hyp.LongFNT& \\
    \quad Hist.gt:& \textcolor{blue}{WADIAK} would basically go and source the ingredients ... \\
    \quad Hist.hyp:& \textcolor{red}{WADIAC} would basically go and source the ingredients ... \\
    \midrule
    Hist.hyp & ... well bessy how are {\color{ogreen} YOU} ... \\
    G.T.& better and not better if {\color{blue} YOU} know what that means \\
    Hyp.FNT& better and not better if {\color{red} YO} know what that means \\
    Hyp.LongFNT& better and not better if {\color{blue} YOU} know what that means  \\
    \midrule
    Hist.hyp & ... getting a meal {\color{ogreen} KIT} every week \\
    G.T.& ... plus the meal {\color{blue} KITS} helped green hone her cooking skills \\
    Hyp.FNT& ... plus the meal {\color{red} KIDS} helped green horn her cooking skills \\
    Hyp.LongFNT& ... plus the meal {\color{blue} KITS} helped green hone her cooking skills \\
    \bottomrule
    \end{tabular}
\end{table}

In order to analyze the effectiveness of the LongFNT model under long-term history, we draw upon the {long-content} attention for token-level integration in $\predv$.
An example of the proposed {long-content} attention during inference is depicted in Figure~\ref{fig:attention}, and the left is the attention in the bottom layer and the right is the attention in the top layer.
As shown in Figure~\ref{fig:att_bottom}, the attention in the bottom layer exhibits a column-based pattern, indicating that specific inputs in the {long-content} text play a crucial role regardless of their position in the input.
Additionally, historical utterance $p-2$ has smaller attention scores compared to utterance $p-1$.
In Figure~\ref{fig:att_top}, the attention in the top layer focuses more on connecting keywords occurred in history and correcting the current word `WADIAK', corresponding to Table~\ref{tab:egs}.
These results demonstrate that the proposed {long-content} attention mechanism can effectively capture long-range context, thereby improving the system performance.

\section{Conclusion}
\label{sec:concl}


In this paper, we introduce two novel approaches, \textbf{LongFNT} and \textbf{SLongFNT}, to incorporate {long-content} information into the factorized neural transducer~(FNT) architecture, and achieve significant improvements in both non-streaming and streaming speech recognition scenarios.

At first, we investigated the effectiveness of incorporating {long-content} history into conformer transducer models and found little improvement.
Subsequently, experiments are conducted based on FNT to explore an efficient network for {long-content} history utilization. We propose LongFNT, using two integration methods for utilizing {long-content} text~(LongFNT-Text) and {long-content} speech~(LongFNT-Speech), and it achieves 19/12\% relative word error rate reduction~(relative WERR) on LibriSpeech/GigaSpeech, outperforming FNT and CT baselines.
We then extended LongFNT to the streaming scenario with SLongFNT, and it achieves 26/17\% relative WERR on LibriSpeech/GigaSpeech, outperforming streaming FNT baselines.
The experiments demonstrate that incorporating {long-content} information can significantly improve ASR performance, and our proposed models offer a promising solution for improving {long-content} speech recognition in real-world scenario with both non-streaming and streaming situations.

\normalem
\bibliographystyle{IEEEtran}
\bibliography{main}

\begin{IEEEbiography}
[{\includegraphics[width=1in,clip]{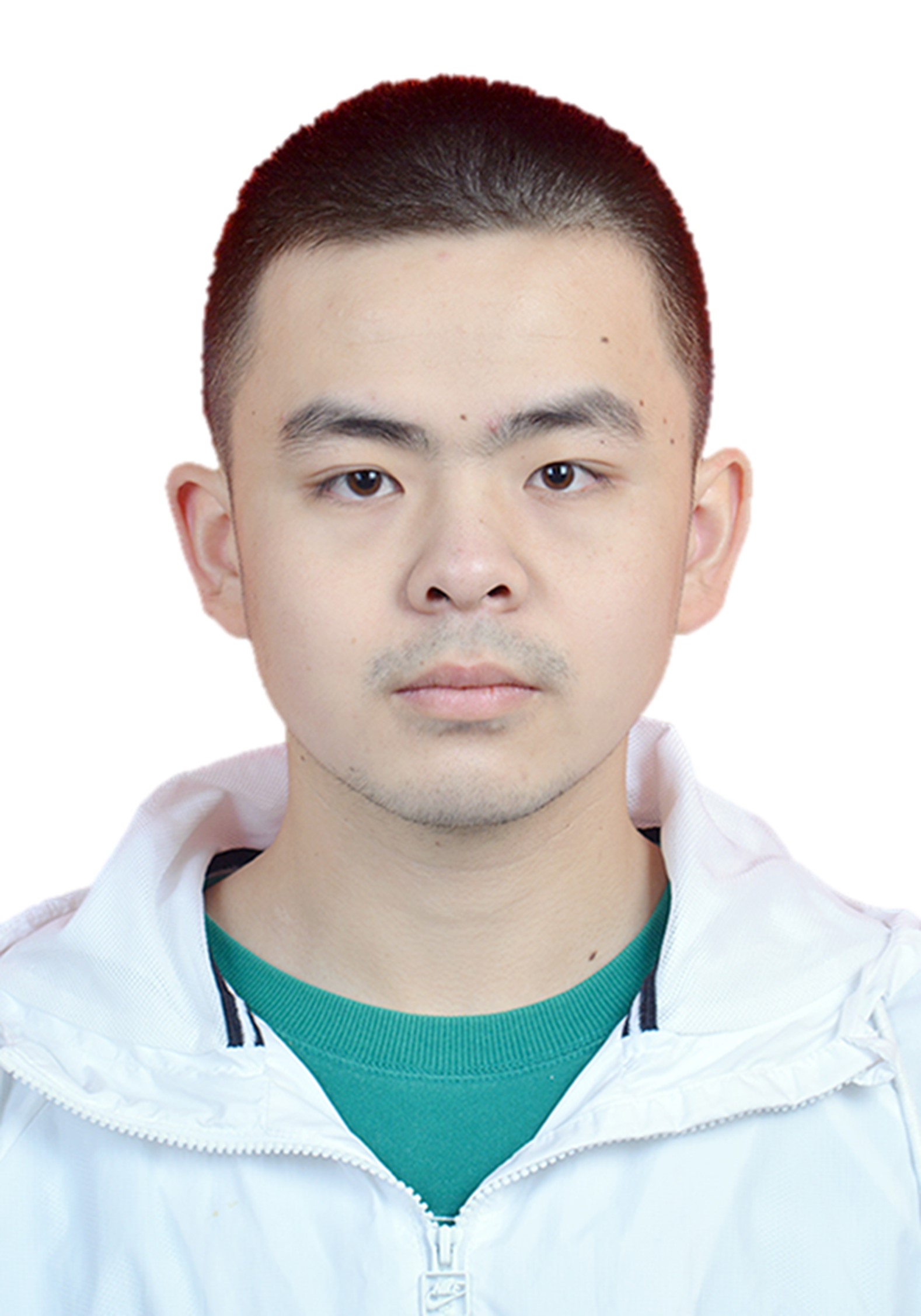}}]
{Xun Gong}~(Student Member, IEEE) received the B.Eng. degree from Zhiyuan College, Shanghai Jiaotong University, Shanghai, China, in 2021. He is currently working toward the Ph.D. degree with the Auditory Cognition and Computational Acoustics Lab, Department of Computer Science and Engineering, Shanghai Jiao Tong University, Shanghai, China, under the supervision of Yanmin Qian. His current research interests include speech recognition and its adaptation.
\end{IEEEbiography}
\begin{IEEEbiography}
[{\includegraphics[width=1in,clip]{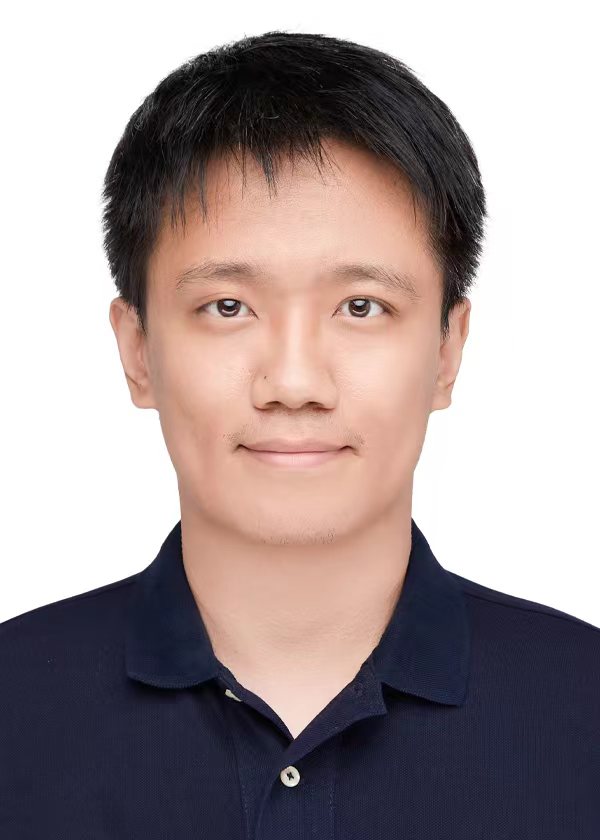}}]
{Yu Wu} is a senior researcher at the Natural Language Computing Group, Microsoft Research Asia (MSRA). He obtained B.S. degree and Ph.D. at Beihang University. His research focuses on end-to-end speech recognition, conversional system, and speech pre-training.
\end{IEEEbiography}
\begin{IEEEbiography}
[{\includegraphics[width=1in,clip]{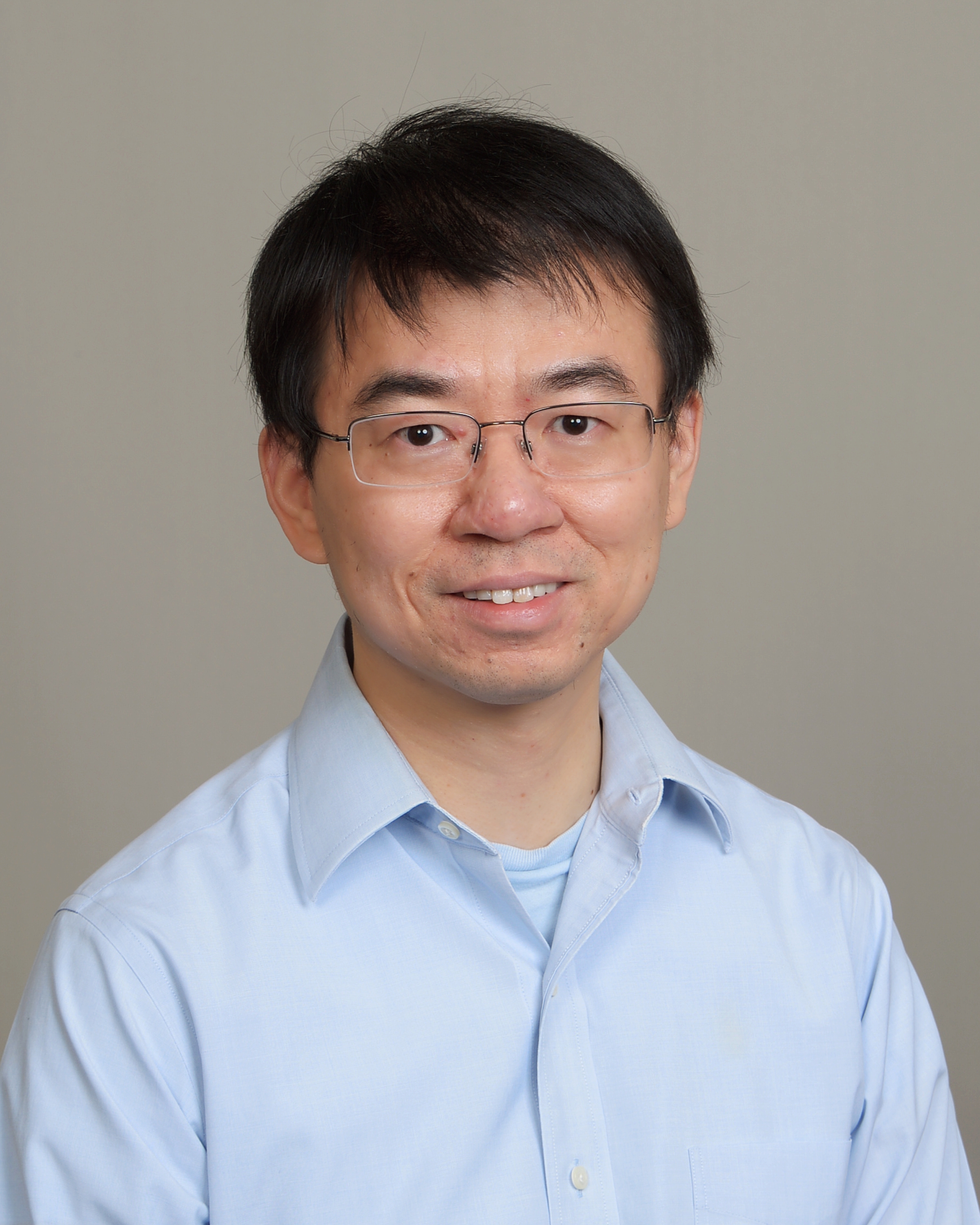}}]
{Jinyu Li} (M'08, SM'21) earned his Ph.D. from Georgia Institute of Technology in 2008. From 2000 to 2003, he was a Researcher in the Intel China Research Center and Research Manager in iFlytek, China. He joined Microsoft in 2008 and now serves as Partner Applied Science Manager, leading a dynamic team dedicated to designing and enhancing speech modeling algorithms and technologies. Their aim is to ensure that Microsoft products maintain cutting-edge quality within the industry. His diverse research areas include end-to-end modeling for speech recognition and speech translation, deep learning, acoustic modeling, and noise robustness. He has been a member of IEEE Speech and Language Processing Technical Committee since 2017. He also served as the associate editor of IEEE/ACM Transactions on Audio, Speech and Language Processing from 2015 to 2020. He was awarded as the Industrial Distinguished Leader at Asia-Pacific Signal and Information Processing Association (APSIPA) in 2021 and APSIPA Sadaoki Furui Prize Paper Award in 2023.
\end{IEEEbiography}
\begin{IEEEbiography}
[{\includegraphics[width=1in,clip]{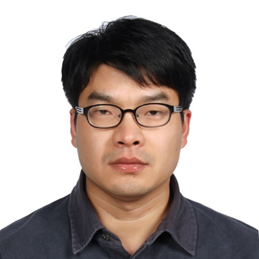}}]
{Shujie Liu} is a principal researcher and research manager in Microsoft Research Asia. He received the Ph.D. degree in computer science from Harbin Institute of Technology, Harbin, China. His research interests include natural language processing and spoken language processing, such as text/speech machine translation (MT), natural language generation (NLG), conversation systems, speech pre-training, text to speech generation (TTS), automatic speech recognition (ASR). 
\end{IEEEbiography}
\begin{IEEEbiography}
[{\includegraphics[width=1in,clip]{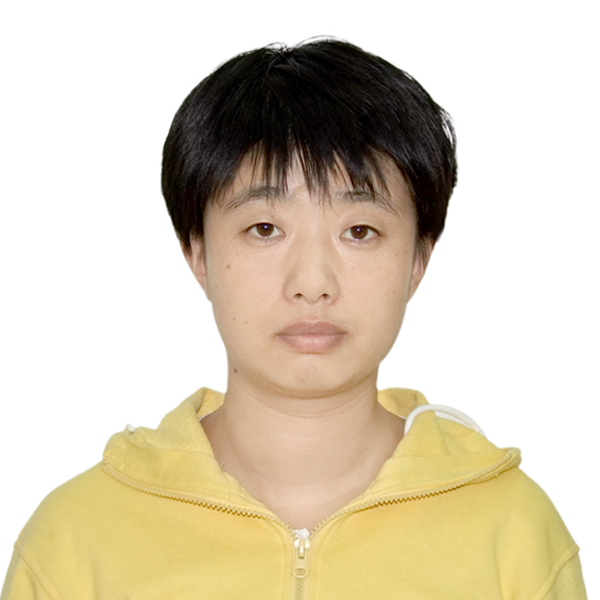}}]
{Rui Zhao} received the Ph.D. degree from Tsinghua University, Beijing, China, in 2005. From 2005 to 2011, she was a Researcher at Toshiba (China) research and development center, where she has been working on Mandarin digit speech recognition, robust speech recognition in car environment, and embedded speech recognition system. Now she is a Principal Applied Scientist at Microsoft, Redmond, WA, USA. Her research interests includes end-to-end speech recognition, end-to- end speech translation.
\end{IEEEbiography}
\begin{IEEEbiography}
[{\includegraphics[width=1in,clip]{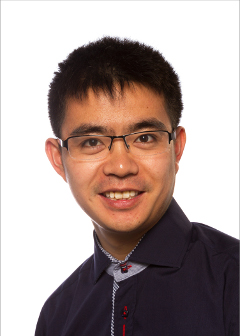}}]
{Xie Chen} is currently a Tenure-Track Associate Professor in the Department of Computer Science and Engineering at Shanghai Jiao Tong University, China. He obtained his Bachelor's degree in the Electronic Engineering department from Xiamen University in 2009, a Master's degree in the Electronic Engineering department from Tsinghua University in 2012, and a Ph.D. degree in the information engineering department at Cambridge University (U.K.) in 2017. Prior to joining SJTU, he worked at Cambridge University as a Research Associate from 2017 to 2018, and in the speech and language research group at Microsoft as a senior and principal researcher from 2018 to 2021. His main research interest lies in deep learning, especially its application to speech processing, including speech recognition and synthesis.
\end{IEEEbiography}
\begin{IEEEbiography}
[{\includegraphics[width=1in,clip]{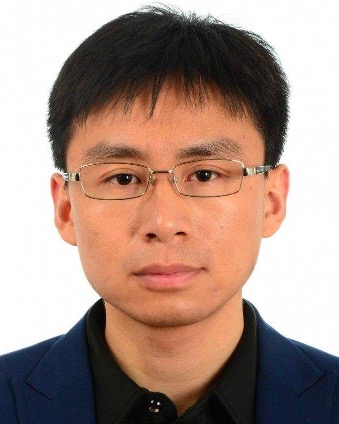}}]
{Yanmin Qian}~(Senior Member, IEEE) received the B.S. degree from the Department of Electronic and Information Engineering, the Huazhong University of Science and Technology, Wuhan, China, in 2007 and the Ph.D. degree from the Department of Electronic Engineering, Tsinghua University, Beijing, China, in 2012. Since 2013, he has been with the Department of Computer Science and Engineering, Shanghai Jiao Tong University, Shanghai, China, where he is currently a Full Professor. From 2015 to 2016, he was an Associate Researcher with the Speech Group, Cambridge University Engineering Department, Cambridge, U.K. He has authored or coauthored more than 200 papers in peer-reviewed journals and conferences on speech and language processing, including T-ASLP, Speech Communication, ICASSP, INTERSPEECH, and ASRU. He has applied for more than 80 Chinese and American patents and was the recipient of the ﬁve championships of international challenges. His research interests include automatic speech recognition and translation, speaker and language recognition, speech separation and enhancement, music generation and understanding, speech emotion perception, multimodal information processing, natural language understanding, and deep learning and multi-media signal processing. He was the recipient of several top academic awards in China, including Chang Jiang Scholars Program of the Ministry of Education, Excellent Youth Fund of the National Natural Science Foundation of China, and the First Prize of Wu Wenjun Artificial Intelligence Science and Technology Award (First Completion). He was also the recipient of several awards from international research committee, including the Best Paper Award in Speech Communication and Best Paper Award from IEEE ASRU in 2019. He is also the Member of IEEE Signal Processing Society Speech and Language Technical Committee.
\end{IEEEbiography}

\end{document}